\documentclass[aps,prx,showpacs,amsmath,amssymb,longbibliography,showkeys,12pt]{revtex4-2}
\usepackage{amsfonts}
\usepackage{amsmath}
\usepackage{amssymb}
\usepackage{graphicx}
\usepackage{epsfig}
\usepackage{exscale}
\usepackage{float}
\usepackage{bm}
\usepackage{suffix}
\usepackage{mathtools}
\usepackage{newunicodechar}
\usepackage{enumitem}
\usepackage{bbding}
\usepackage{tikz}
\usepackage{multirow}
\usepackage{booktabs}
\usepackage{nameref}
\usepackage{xcolor}
\usepackage[colorlinks=true,linkcolor=blue]{hyperref}
\makeatletter
\def\p@subsection{}
\def\p@subsubsection{}
\makeatother
\DeclarePairedDelimiterX\MeijerM[3]{\lparen}{\rparen}
{\begin{smallmatrix}#1 \\ #2\end{smallmatrix}\delimsize\vert\,#3}

\newif\ifmarkup
\markupfalse
\ifmarkup
  \newcommand{\rev}[1]{{\color{blue}#1}}
  \newenvironment{revblock}{\color{blue}}{}
  \newcommand{\revcolor}{\color{blue}}
\else
  \newcommand{\rev}[1]{#1}
  \newenvironment{revblock}{}{}
  \newcommand{\revcolor}{}
\fi

\begin{document}
\title{Instanton versus Renormalon Dominance in the Large-Order Behavior of
\texorpdfstring{\(O(N)\)}{O(N)}-Symmetric \texorpdfstring{\(\phi^4\)}{phi4} Theory}
\author{Abouzeid M. Shalaby}
\affiliation{Department of Physics and Materials Sciences, College of Arts and Sciences,
Qatar University, Al Tarfa, Doha 2713, Qatar}
\keywords{Instanton large-order behavior, Renormalon large-order behavior,
Hypergeometric approximant, \texorpdfstring{\(O(N)\)}{O(N)}-symmetric
\texorpdfstring{\(\phi^4\)}{phi4} theory}

\begin{abstract}
The eight-loop field anomalous dimension \(\gamma_\phi\) is the highest (\rev{exact})
perturbative order available for the renormalization-group functions of the
\(O(N)\)-symmetric \(\phi^4\) theory. We use it to test whether the large-order
behavior, \(c_k \sim \gamma\,k!\,\sigma^k k^\delta\), is of instanton or of
renormalon type. Motivated by the comparatively rapid convergence of \(\sigma\),
we introduce a \rev{two-step} \(\sigma\)-test. A Gevrey-1 hypergeometric
approximant is built from the seven-loop series with \(\sigma\) fixed to its
instanton or renormalon value\rev{, which accelerates the convergence of
\(\delta\)}. The approximant then predicts the known eight-loop coefficient and the
exponent \(\delta\)\rev{, and the two predictions are judged together. A control
computation on the anharmonic oscillator shows why: the next coefficient is
reproduced for a wide range of \(\sigma\), whereas \(\delta\) is sensitive to it}.
The combined evidence favors instanton-type behavior for all \(N\) considered,
most clearly for large \(N\); for small \(N\) the decision rests on \(\delta\).
\rev{The value of \(\sigma\) required to reproduce the instanton exponent lies
within about \(6\%\) of \(\sigma_{\rm inst}=-1\) for every \(N\), whereas the
renormalon exponent cannot be reproduced at all for \(N=1,3\) and requires
\(\sigma\) to deviate from \(\sigma_{\rm ren}=(N+8)/6\) by \(17\%\) (\(N=5\)) to a
factor of about 35 (\(N=20\)).}
\end{abstract}
\maketitle

\section{Introduction}

Perturbative expansions in quantum field theory are, in general, divergent
asymptotic series rather than convergent power series. Their large-order
behavior is physically significant because it encodes information about the
non-perturbative structure of the theory through the singularities of the Borel
transform. In a renormalizable theory two mechanisms are usually identified as
sources of the factorial growth of the perturbative coefficients: semiclassical
saddle points, commonly referred to as instantons, and diagrammatic renormalon
singularities associated with particular classes of Feynman graphs and with the
renormalization-group structure of the theory
\cite{Lipatov1977,tHooft1979,Parisi1978,Parisi1979,Beneke1999,ZinnJustin1997,LGLZJ1990}.

If a perturbative quantity admits an expansion of the form
\begin{equation}
F(g)=\sum_{k=0}^{\infty} c_k g^k ,
\end{equation}
then, for a Gevrey-1 asymptotic series, its leading large-order behavior is
expected to take the form
\begin{equation}
c_k \sim \gamma\, k!\,\sigma^k k^{\delta}
\left[1+\mathcal{O}\left(\frac{1}{k}\right)\right],
\qquad k\rightarrow \infty .
\label{eq:large-order-generic}
\end{equation}
Here \(1/\sigma\) determines the location of the leading Borel singularity,
\(\delta\) characterizes the nature of that singularity, and \(\gamma\) measures
its strength \cite{ZinnJustin1997,LGLZJ1990,Renormalon22}. The question of
whether instantons or renormalons dominate the large-order behavior therefore
reduces, in practice, to identifying the leading Borel singularity and its
dependence on the parameters of the theory.

The \(O(N)\)-symmetric \(\phi^4\) theory is a particularly important testing
ground for this question. On the one hand, the model is central to the
field-theoretic description of critical phenomena and of the Wilson--Fisher
fixed point. On the other hand, it is one of the few interacting quantum field
theories for which relatively high-order renormalization-group data are
available. Following the classic five-loop computations
\cite{Gorishnii1983,Gorishnii1984,Kleinert1991}, the beta function and the
anomalous dimensions of the \(O(N)\)-symmetric \(\phi^4\) theory in the
minimal-subtraction scheme have been computed to six and seven loops
\cite{Borel-6L,Schnetz2018,Schnetz2023}.  \rev{However}, the eight-loop coefficient of the field anomalous dimension
\rev{has since been computed by Schnetz~\cite{Schnetz2023}; the results are
also distributed with his program package \textsc{HyperlogProcedures}~\cite{Schnetz-06-23}}. Since these results carry the full
\(N\)-dependence, they provide a nontrivial opportunity to test the large-order
behavior of the theory and to examine whether the available coefficients favor
instanton or renormalon dominance.

\begin{revblock}
Eight loops is the highest order at which the \emph{complete}
renormalization-group functions are known. Considerably more is known about the
primitive (subdivergence-free) contribution to the beta function, which is
conjectured to dominate its large-order behavior in the minimal-subtraction
scheme: the corresponding Feynman periods have been evaluated numerically up to
17--18 loops \cite{Balduf2023,BaldufShaban2024,BaldufThurigen2024,BorinskyFavorito2025},
and the primitive beta function with its full \(O(N)\) dependence up to 17 loops
\cite{Balduf2023,BaldufThurigen2024}.
A central lesson of these studies is that the asymptotic regime is approached
slowly. In the zero-dimensional \(O(N)\) model the leading growth rate becomes
visible only above roughly 25 loops, and lower-order data suggest a wrong
asymptotics; in four dimensions the growth rate of the primitive beta function
has not reached its asymptotic value even at 17 loops, although its
\(N\)-dependence is close to the asymptotic form already at about ten loops
\cite{BaldufThurigen2024}. Any statement about large-order behavior extracted
from seven or eight loops must be read in this light; we return to this point
in Sec.~\ref{sec:asymptotics}.
\end{revblock}

For the beta-function coefficients,
\begin{equation}
\beta(g;N)=\sum_{k\geq 0}\beta_k(N)g^k ,
\end{equation}
the instanton analysis predicts the leading large-order behavior
\cite{Lipatov1977,Brezin1977,McKaneWallace1978,McKane2019,Renormalon22}
\begin{equation}
\beta_k^{\mathrm{inst}}(N)
\sim
\gamma_{\mathrm{inst}}\,
k!\,\sigma_{\mathrm{inst}}^{\,k} k^{\delta_{\mathrm{inst}}}
\left(1+\mathcal{O}\left(\frac{1}{k}\right)\right),
\qquad k\rightarrow \infty ,
\label{eq:instanton-beta}
\end{equation}
which, in the notation of Eq.~\eqref{eq:large-order-generic}, corresponds to
\begin{equation}
\sigma_{\mathrm{inst}}=-1,
\qquad
\delta_{\mathrm{inst}}(N)=3+\frac{N}{2}.
\label{instanton-ab}
\end{equation}

The renormalon analysis
\rev{\cite{Parisi1978,Parisi1979,ren1981,ren1987,ren1988,Renormalon22}} leads to a
qualitatively different prediction. In terms of the first two nontrivial
beta-function coefficients \rev{\cite{Kleinert-Borel}},
\begin{equation}
\beta_2(N)=\frac{N+8}{3},
\qquad
\beta_3(N)=-\frac{3N+14}{3},
\end{equation}
the renormalon large-order behavior is
\begin{equation}
\beta_k^{\mathrm{ren}}(N)
\sim
\gamma_{\mathrm{ren}}\,
k!\,\sigma_{\mathrm{ren}}^{\,k} k^{\delta_{\mathrm{ren}}}
\left(1+\mathcal{O}\left(\frac{1}{k}\right)\right),
\qquad k\rightarrow \infty ,
\label{eq:renormalon-beta}
\end{equation}
with the parameters quoted in Ref.~\cite{Renormalon22},
\begin{equation}
\sigma_{\mathrm{ren}}(N)=\frac{N+8}{6},
\qquad
\delta_{\mathrm{ren}}(N)=
\frac{6(3N+14)}{(N+8)^2}.
\label{renormalon-ab}
\end{equation}
The \(N\)-dependence in Eqs.~\eqref{instanton-ab} and \eqref{renormalon-ab} is
thus very different. This distinction is crucial: it allows the available
high-loop data to be used not only to test the existence of factorial growth, but
also to discriminate between the two mechanisms.

Recent analysis of the exact six- and seven-loop \(\overline{\mathrm{MS}}\)
results, supplemented by estimates of higher-order primitive-graph contributions,
indicates that the large-order behavior of the \(O(N)\)-symmetric \(\phi^4\) beta
function is more consistent with the instanton prediction than with the
renormalon one \cite{Borel-6L,Schnetz2017,Renormalon22}. In particular, ratios of
successive coefficients normalized with the instanton values of \(\sigma\) and
\(\delta\) approach the expected asymptotic behavior more naturally than the
corresponding renormalon-normalized ratios. The distinction is especially clear
in the sign of \(\sigma\) and in the \(N\)-dependence of the subleading exponent
\(\delta\). However, Fig.~3 of Ref.~\cite{Renormalon22}, which tests the
large-order parameter \(\delta\), can barely differentiate between the two
scenarios.

This observation is consistent with the long-standing suggestion that
renormalization-group functions in the \(\overline{\mathrm{MS}}\) scheme may be
insensitive to renormalons \cite{BergereDavid1984,McKane2019,Renormalon22}. If
true, this would imply that the visibility of renormalons is not an intrinsic
property of the theory alone but depends on the renormalization scheme and on
the observable under consideration. The issue is not yet settled, however. The
available exact perturbative data are impressive but still limited, and the
approach to the true asymptotic regime can be slow, as is already known from
simpler quantum-mechanical analogues such as the quartic anharmonic oscillator
\cite{entire,Renormalon22} \rev{and, for the primitive beta function, from the
high-loop studies cited above \cite{BaldufThurigen2024}}. Thus, while present
evidence favors instanton dominance in the large-order behavior of the
\(\overline{\mathrm{MS}}\)-scheme \(O(N)\)-symmetric \(\phi^4\)
renormalization-group functions, a complete understanding of the possible
suppression or absence of renormalons remains an open problem.

In view of the recent eight-loop result for the field anomalous dimension
\(\gamma_{\phi}\), and motivated by the observation in our previous work
\cite{entire} that the parameter \(\sigma\) converges to its exact value more
rapidly than the parameter \(\delta\), we propose in this work what we call the
\(\sigma\)-test. The test is based on the hypergeometric resummation algorithm we
developed in Refs.~\cite{eps7,abohop,abo-precize}, which can both predict and
accommodate the large-order parameters of a divergent perturbative series. It
proceeds by constructing a hypergeometric parametrization of the series from the
seven-loop data together with a prescribed value of \(\sigma\); the resulting
approximant then predicts the next known coefficient, namely the eight-loop one,
and the large-order exponent \(\delta\). The procedure is carried out with the
instanton and with the renormalon value of \(\sigma\).

\begin{revblock}
Two remarks delimit what such a test can establish. First, the coefficients of
\(\gamma_\phi\) alternate in sign for every \(N\), which by itself indicates a
leading Borel singularity on the negative real axis, \(\sigma<0\). A test that
only confirmed the sign of \(\sigma\) would therefore add little. The relevant
question is whether the data are consistent with the instanton
\emph{magnitude} \(|\sigma|=1\) and, simultaneously, with the instanton exponent
\(\delta\) and its \(N\)-dependence, rather than, for instance, with
\(\sigma=-(N+8)/6\). We address this by separating the roles of the sign and the
magnitude of \(\sigma\), and by a joint test in which the value of \(\sigma\)
required to reproduce the instanton exponent is determined from the data.
Second, the test consists of two steps, and both of its outputs must be read
together. In the first step the exact \(\sigma\) is used to accelerate the
convergence of the non-perturbative exponent \(\delta\), which a finite-order
approximant with \(\sigma\) left free reproduces poorly. In the second step the
scenario is judged by the prediction of the unused eight-loop coefficient
\emph{and} by the predicted \(\delta\). A test of this kind is meaningful only if a
wrong input \(\sigma\) produces a noticeably worse result in a case where the answer
is known. We therefore perform control computations with wrong values of \(\sigma\)
for the quartic anharmonic oscillator. They show that an accurate next
coefficient is a necessary but not a sufficient condition, since it is obtained
for a wide range of \(\sigma\), while the predicted exponent is highly sensitive to
\(\sigma\). A scenario is therefore regarded as favored only if its
parametrization satisfies both criteria.
\end{revblock}

The remainder of the paper is organized as follows. Section~\ref{sec:method}
collects the exact large-order parameters relevant to \(\gamma_\phi\),
\rev{gives a step-by-step derivation of the renormalon parameters,} and
summarizes the hypergeometric algorithm. Section~\ref{sec:AHO} benchmarks the
\(\sigma\)-test on the quartic anharmonic oscillator, where all three large-order
parameters are known exactly, \rev{including control computations with wrong
values of \(\sigma\)}. Section~\ref{sec:smallN} applies the test to the
eight-loop \(\gamma_\phi\) series for \(N=1,3,5\), and Sec.~\ref{sec:largeN} to
\(N=10,15,20\). \rev{Section~\ref{sec:joint} separates the sign and the magnitude
of \(\sigma\) and presents the joint \((\sigma,\delta)\) test.
Section~\ref{sec:asymptotics} discusses the results in the light of the
high-order studies of primitive graphs.} Section~\ref{sec:conclusions} summarizes
the results.

\section{Large-order parameters of \texorpdfstring{\(\gamma_\phi\)}{gamma phi}
and the \texorpdfstring{\(\sigma\)}{sigma}-test}
\label{sec:method}

\subsection{Exact large-order parameters}
\label{sec:exact}

\begin{revblock}
We write the running coupling of the four-dimensional \(O(N)\) \(\phi^4\) theory
as
\begin{equation}
\mu\frac{dg}{d\mu}=\beta(g)=\beta_2 g^2+\beta_3 g^3+\dots,
\qquad
\beta_2=\frac{N+8}{3},\quad \beta_3=-\frac{3N+14}{3},
\label{eq:beta-2loop}
\end{equation}
and derive the renormalon parameters directly in the minimal-subtraction scheme,
the scheme in which both \(\gamma_\phi\) and the instanton parameters of
Eq.~\eqref{eq:inst-gammaphi} are given. We follow the bubble-chain analysis of
Ref.~\cite{Suzuki1998}, in four steps.

\emph{Step 1: MS coupling and the bubble chain.} In \(d=4-2\epsilon\) the MS
coupling is defined by \(u_0=g\,Z_g(g,\epsilon)\), with \(u_0\) the dimensionless
bare coupling and \(Z_g\) a pure pole series \cite{tHooftDimReg1973}. Summing the
chains of one-loop bubbles in the four-point function and requiring finiteness
gives \(Z_g=1/(1-\beta_2g/2\epsilon)\), \(\beta(g)=-2\epsilon g+\beta_2g^2\), and
the renormalized chain coupling at Euclidean momentum \(q\),
\begin{equation}
\Delta(q^2)=\frac{g}{1-g\,\Pi(q^2)},\qquad
\Pi(q^2)=\frac{\beta_2}{2}\Bigl[\ln\frac{q^2}{\mu^2}-c_1\Bigr],
\label{eq:chainMS}
\end{equation}
i.e.\ the one-loop running coupling. Its Borel transform with respect to \(g\)
is \(e^{\Pi t}\propto(q^2/\mu^2)^{\beta_2t/2}\).

\emph{Step 2: location of the singularity.} Let \(\Delta\) be inserted as the
dressed line of a skeleton graph of a dimensionless quantity. The ultraviolet
region that reproduces a dimension-six operator, the leading one
\cite{Parisi1978,Parisi1979,BergereDavid1984}, contributes
\begin{equation}
F(g)=\int_{\mu^2}^{\infty}\frac{dq^2}{q^2}\,\frac{\mu^2}{q^2}\,\Delta(q^2).
\label{eq:FMS}
\end{equation}
Its Borel transform is proportional to \(1/(1-\beta_2t/2)\), a pole at
\(t=2/\beta_2\), so that
\begin{equation}
\sigma_{\rm ren}=\frac{\beta_2}{2}=\frac{N+8}{6}.
\label{eq:match}
\end{equation}
At the level of the coefficients, \(\int_0^\infty dx\,e^{-x}x^k=k!\) gives
\(c_{k+1}\simeq k!\,\sigma^k\), i.e.\ \(c_m\sim\sigma^m\Gamma(m+1+b)\) with
\(b=-1\).

\emph{Step 3: exponent.} Two effects modify the pure factorial. First, the
two-loop running, \(d\tilde g/dx=\tfrac12(\beta_2\tilde g^2+\beta_3\tilde g^3)\),
multiplies the coefficients by \(k^{2\beta_3/\beta_2^2}\), with the signed
\(\beta_3\). Second, the Wilson coefficient of the operator \((\phi^2)^3\) runs
with its anomalous dimension, which multiplies the integrand of Eq.~\eqref{eq:FMS}
by \((g/\Delta)^{\kappa}\), \(\kappa=\gamma_6^{(1)}/\beta_2\). The one-loop
anomalous dimension of \((\phi^2)^k\) is \(\gamma_{(\phi^2)^k}=k(N+6k-4)g/3\)
\cite{KehreinWegnerPismak1993}; as a check, \(k=1\) reproduces the mass anomalous dimension
\((N+2)g/3\) and \(k=2\) gives \(2\beta_2g\), as required for the operator
conjugate to \(g\). For \(k=3\), \(\gamma_6^{(1)}=N+14\) and
\(\kappa=3(N+14)/(N+8)\). In the one-loop chain the factor turns the integrand
into \(g(1-x)^{\kappa-1}\), whose coefficients carry
\(\Gamma(n+1-\kappa)\), so that \(b\) is lowered by \(\kappa\). The renormalized
coupling defined by \(F\) therefore has coefficients
\(r_k\sim\sigma^k\,\Gamma(k+1+b_r)\) with
\begin{align}
b_r&=\frac{2\beta_3}{\beta_2^2}-1-\kappa=\nu-1,\nonumber\\
\nu(N)&=-\frac{3(N^2+28N+140)}{(N+8)^2}.
\label{eq:nu}
\end{align}
In Parisi's notation the coefficients behave as \(\Gamma(k-\alpha_6)\) with
\(\alpha_6=-\nu\) \cite{Parisi1978}. For \(N\to\infty\), \(\nu\to-3\), in
agreement with the large-\(N\) value \(\alpha=3\) of
Refs.~\cite{Parisi1978,Parisi1979}. We have confirmed both terms of \(b_r\) by
expanding the exact two-loop chain to high order.

\emph{Step 4: transfer to \(\gamma_\phi\).} Let \(\bar g=g+r(g)\) be the coupling
defined by \(F\). The field anomalous dimension transforms as a function,
\(\bar\gamma_\phi(\bar g)=\gamma_\phi(g(\bar g))\), so that, to first order in
\(r\), its coefficients change by
\(\delta\gamma_k=-\sum_j j\gamma_j\,r_{k-j+1}\simeq-2\gamma_2\,r_{k-1}\), because
\(\gamma_\phi=\gamma_2g^2+\dots\) starts at order \(g^2\). The shift of index
lowers the exponent by one unit, \(b_r-1\). We have checked this with the exact
\(\gamma_2,\dots,\gamma_8\): for \(N=1\) the ratio
\(\delta\gamma_k/(-2\gamma_2r_{k-1})\) is \(0.961\), \(0.984\), \(0.992\), \(0.996\),
\(0.998\) at \(k=20,50,100,200,400\). Hence \(\delta_{\gamma_\phi}=\nu-2\), and
\begin{equation}
\delta^{\gamma_\phi}_{\rm ren}=-\frac{5N^2+116N+548}{(N+8)^2},
\qquad
\sigma^{\gamma_\phi}_{\rm ren}=\frac{N+8}{6}.
\label{eq:ren-gammaphi}
\end{equation}

\emph{What the reference means.} The relation between \(\gamma_\phi\) and
\(\bar\gamma_\phi\) fixes only their difference; it does not say which of the two
series carries the renormalon. If \(\bar\gamma_\phi\) is free
of factorial growth, the MS coefficients of \(\gamma_\phi\) carry the renormalon
with the parameters of Eq.~\eqref{eq:ren-gammaphi}. If instead the MS RG functions
are renormalon free, as conjectured in
Refs.~\cite{BergereDavid1984,McKane2019,Renormalon22}, the MS series shows no
renormalon growth at all. Both the instanton and the renormalon reference values
therefore refer to the \(\overline{\rm MS}\) series of \(\gamma_\phi\), the
renormalon values being those that a renormalon would have \emph{if} it is
present there. The \(\sigma\)-test examines whether the data show any sign of it.

\end{revblock}

For the instanton large-order behavior of \(\gamma_\phi\) one has
\cite{Kleinert-Borel}
\begin{equation}
\delta^{\gamma_\phi}_{\rm inst}=2+\frac{N}{2},
\qquad
\sigma^{\gamma_\phi}_{\rm inst}=-1.
\label{eq:inst-gammaphi}
\end{equation}

A remark is needed before these parameters are compared with the output of
the approximants. The series that is actually parametrized below is not
\(\gamma_\phi\) itself but \(\gamma_\phi/g^2\) with a further overall factor
\(g\) extracted from the hypergeometric part [see
Eq.~\eqref{eq:param-gammaphi}]. Each extracted power of \(g\) shifts the index
of the coefficients by one unit and hence shifts \(\delta\)
by \(+1\), so the reference values against which the predicted exponent must be
compared are \(\delta+3\). The exact parameters used in the tests are collected
in Table~\ref{tab:exact}.

\begin{table}[H]
\centering
\caption{Exact large-order parameters used as reference values in the
\(\sigma\)-test for the parametrized series (\(\gamma_\phi/g^2\) with one further
factor of \(g\) extracted). The shifted exponents are
\(\delta_{\rm inst}=5+N/2\) and \(\delta_{\rm ren}=3-(5N^2+116N+548)/(N+8)^2\).}
\label{tab:exact}
\begin{tabular}{c c c c c}
\toprule
\(N\) & \(\sigma_{\rm inst}\) & \(\sigma_{\rm ren}=(N+8)/6\) &
\(\delta_{\rm inst}\) & \(\delta_{\rm ren}\) \\
\midrule
1  & \(-1\) & \(3/2\)   & 5.5  & \(-5.2593\) \\
3  & \(-1\) & \(11/6\)  & 6.5  & \(-4.7769\) \\
5  & \(-1\) & \(13/6\)  & 7.5  & \(-4.4142\) \\
10 & \(-1\) & \(3\)     & 10   & \(-3.8148\) \\
15 & \(-1\) & \(23/6\)  & 12.5 & \(-3.4518\) \\
20 & \(-1\) & \(14/3\)  & 15   & \(-3.2092\) \\
\bottomrule
\end{tabular}
\end{table}

\subsection{Hypergeometric parametrization}

The hypergeometric--Meijer resummation algorithm developed in
Refs.~\cite{eps7,abo-precize,Abo-large} is well suited to this test, because it
can predict the three parameters defining the large-order behavior of a divergent
series from a finite number of perturbative coefficients. For a Gevrey-1 series
\begin{equation}
Q(x) \approx \sum_{i=0}^{M} c_i x^i ,
\label{pertQ}
\end{equation}
with the large-order behavior
\begin{equation}
c_n =
\gamma\, n! \,\sigma^n\, n^{\delta}
\left(1+\mathcal{O}\left(\frac{1}{n}\right)\right),
\qquad n\rightarrow \infty ,
\label{large-order}
\end{equation}
the series is parametrized by a generalized hypergeometric function,
\begin{equation}
Q(x)
\approx
c_0\, {}_{p}F_{p-2}
\left(
a_1,a_2,\ldots,a_p;
b_1,b_2,\ldots,b_{p-2};
\sigma x
\right),
\label{hyper-param}
\end{equation}
where \(M=2p-1\). When \(\sigma\) is known it can be incorporated directly into
the approximant, which accelerates convergence; in that case only \(M=2p-2\)
perturbative coefficients are needed.

As shown in Refs.~\cite{eps7,abo-precize,Abo-large}, the approximant in
Eq.~\eqref{hyper-param} has the same large-order form as
Eq.~\eqref{large-order}, with
\begin{equation}
\gamma
=
c_0\,
\frac{\prod_{k=1}^{p-2}\Gamma(b_k)}
{\prod_{k=1}^{p}\Gamma(a_k)}
\end{equation}
and
\begin{equation}
\sum_{i=1}^{p} a_i
-
\sum_{i=1}^{p-2} b_i
-2
=
\delta .
\label{constr}
\end{equation}
The analytic continuation of the hypergeometric function follows from the
Mellin--Barnes representation \cite{Analytic2016},
\begin{equation}
{}_{p}F_{q}
(a_1,\ldots,a_p;b_1,\ldots,b_q;z)
=
\frac{\prod_{k=1}^{q}\Gamma(b_k)}
{\prod_{j=1}^{p}\Gamma(a_j)}
\frac{1}{2\pi i}
\int_C
\frac{
\Gamma(s)\prod_{j=1}^{p}\Gamma(a_j-s)
}{
\prod_{k=1}^{q}\Gamma(b_k-s)
}
(-z)^{-s}\,ds ,
\label{hyp-G-C2}
\end{equation}
or equivalently from the continuation formula \cite{Analytic2016}
\begin{align}
&  \,_{p}F_{q}(a_{1},\ldots,a_{p};b_{1},\ldots,b_{q};\sigma z)\nonumber\\
&  =\frac{\prod_{i=1}^{q}\Gamma\left(  b_{i}\right)  }{\prod_{i=1}^{p}
\Gamma\left(  a_{i}\right)  }\sum_{k=1}^{p}(-\sigma z)^{-a_{k}}
\frac{\Gamma\left(  a_{k}\right)  }{\prod_{j=1}^{q}\Gamma\left(  b_{j}
-a_{k}\right)  }\prod_{j=1,j\neq k}^{p}\Gamma\left(  a_{j}-a_{k}\right)
\times\nonumber\\
&  \ \,_{q+1}F_{p-1}\left(  a_{k},a_{k}-b_{1}+1,\ldots,a_{k}-b_{q}
+1;-a_{1}+a_{k}+1,\underbrace{\ldots}_{\ast},-a_{p}+a_{k}+1;\frac{(-1)^{p-q+1}
}{\sigma z}\right),  \label{entire}
\end{align}
where the asterisk indicates that terms of the form \(1-a_{i}+a_{j}\) with
\(i=j\) are excluded.

The algorithm can therefore predict all parameters defining the large-order
behavior of a divergent series from a finite number of perturbative terms.
Conversely, it is well known that feeding non-perturbative large-order
information into a resummation algorithm accelerates its convergence. We
therefore use \(\sigma\) as an input in the construction of the approximant and
test its prediction of both the next unused perturbative coefficient and the
exponent \(\delta\). Of the parameters appearing in Eq.~(\ref{large-order}),
only \(\sigma\) and \(\delta\) are relevant to the present analysis of
\(\gamma_\phi\): \(\sigma\) distinguishes the instanton from the renormalon
scenario and is supplied as input to the hypergeometric approximant, while
\(\delta\) is predicted by the approximant and compared with its known value in
each scenario. The overall normalization constant will not be considered.

\begin{revblock}
For the \({}_3F_1\) approximants used below, the dependence of the predicted
exponent on the input \(\sigma\) can be made explicit. As shown in
Sec.~\ref{sec:AHO}, the parameters follow from the linear system
\eqref{eq:linear-system} for the coefficients \(f_0,f_1,f_2\) and \(e_1\), with
\(e_2=1\) and \(f_3=\sigma\). Denote by
\(r_n\equiv h_n/h_{n-1}\) (\(n=1,\ldots,4\)) the ratios of successive
coefficients of the series to be approximated; they are fixed by the
perturbative input alone and do not depend on \(\sigma\). Moving the known
terms to the right-hand side, Eq.~\eqref{eq:linear-system} becomes
\begin{equation}
f_0+f_1\,n+f_2\,n^2-e_1\,n\,r_n=n^2 r_n-\sigma\,n^3,\qquad n=1,\ldots,4 .
\label{eq:lin-sigma}
\end{equation}
In matrix form, \(M\mathbf{x}=\mathbf{u}-\sigma\,\mathbf{v}\), with
\(\mathbf{x}=(f_0,f_1,f_2,e_1)^{T}\), the \(n\)-th row of \(M\) equal to
\((1,\,n,\,n^2,\,-n\,r_n)\), \(u_n=n^2 r_n\) and \(v_n=n^3\). Neither \(M\) nor
\(\mathbf{u}\) and \(\mathbf{v}\) contain \(\sigma\); the input \(\sigma\)
appears only as the coefficient of \(\mathbf{v}\) on the right-hand side. The
solution is therefore affine in \(\sigma\),
\begin{equation}
\mathbf{x}(\sigma)=M^{-1}\mathbf{u}-\sigma\,M^{-1}\mathbf{v}
\equiv\mathbf{x}^{(0)}-\sigma\,\mathbf{x}^{(1)},
\label{eq:affine}
\end{equation}
and in particular \(f_2=f_2^{(0)}-\sigma f_2^{(1)}\) and
\(e_1=e_1^{(0)}-\sigma e_1^{(1)}\), where the four constants
\(f_2^{(0)},f_2^{(1)},e_1^{(0)},e_1^{(1)}\) follow from Cramer's rule as ratios
of determinants built from \(r_1,\ldots,r_4\).

The exponent is obtained from these two coefficients. Expanding the numerator
and denominator of \(h_n/h_{n-1}\) in powers of \(n\),
\begin{align}
\sigma\prod_{i=1}^{3}(a_i+n-1)&=\sigma n^3+\sigma\Big(\sum_{i=1}^{3}a_i-3\Big)n^2
+\cdots,\nonumber\\
n\,(b_1+n-1)&=n^2+(b_1-1)\,n ,
\end{align}
and comparing with \(\sum_i f_i n^i\) and \(\sum_j e_j n^j\) gives
\(f_2=\sigma\big(\sum_i a_i-3\big)\) and \(e_1=b_1-1\), i.e.\
\(\sum_i a_i=3+f_2/\sigma\) and \(b_1=1+e_1\). For \(p=3\), \(q=1\),
Eq.~\eqref{constr} reads \(\delta=\sum_i a_i-b_1-2\), so that
\begin{equation}
\delta=\frac{f_2}{\sigma}-e_1 .
\end{equation}
Inserting the affine forms of \(f_2\) and \(e_1\) yields the exact form
\begin{equation}
\delta_{\rm pred}(\sigma)=\frac{A}{\sigma}+B-C\,\sigma ,
\label{eq:delta-sigma}
\end{equation}
with
\begin{equation}
A=f_2^{(0)},\qquad B=-\big(f_2^{(1)}+e_1^{(0)}\big),\qquad C=-e_1^{(1)} .
\label{eq:ABC}
\end{equation}
The constants \(A,B,C\) are thus fixed by the input coefficients alone. For the
anharmonic oscillator of Sec.~\ref{sec:AHO} one finds \(A=-80.48\),
\(B=-45.50\), \(C=6.625\), which at \(\sigma=-3\) reproduces
\(\delta_{\rm pred}=1.20\); for \(\gamma_\phi\) at \(N=1\) one finds
\(A=-13.10\), \(B=-9.63\), \(C=1.50\), and at \(N=20\)
\(A=-19.73\), \(B=-5.98\), \(C=-0.018\). We have checked that
Eq.~\eqref{eq:delta-sigma} with these constants agrees with the direct solution
of the linear system.

Equation \eqref{eq:delta-sigma} shows directly why \(\delta_{\rm pred}\) is so
sensitive to \(\sigma\): the term \(A/\sigma\) diverges as \(\sigma\to0\), the
term \(C\sigma\) grows linearly for large \(|\sigma|\), and a change of the input
\(\sigma\) by \(\Delta\sigma\) shifts the exponent by
\(\Delta\delta\simeq-(A/\sigma^2+C)\,\Delta\sigma\), which is of order
\(|A|\) for \(|\sigma|\approx1\). It also makes the joint test of
Sec.~\ref{sec:joint} transparent: requiring
\(\delta_{\rm pred}(\sigma)=\delta_{\rm ref}\) is the quadratic equation
\begin{equation}
C\sigma^2+(\delta_{\rm ref}-B)\,\sigma-A=0
\label{eq:joint-quadratic}
\end{equation}
for \(\sigma\), which has at most two real roots. The same affine
structure fixes the \(\sigma\)-dependence of the predicted next coefficient,
which is discussed in Sec.~\ref{sec:joint}.

The \(\sigma\)-test thus consists of two steps.
\begin{enumerate}[label=(\roman*),leftmargin=*,itemsep=1pt]
\item \emph{Acceleration.} The exact \(\sigma\) of a scenario is supplied as input.
This removes one unknown from the linear system, so that the finite set of
coefficients is spent on the remaining parameters, and the predicted exponent
converges much faster than when \(\sigma\) must be predicted as well. For the
oscillator, \(\delta_{\rm pred}\) improves from \(2.76\) to \(1.20\) (exact
\(0.5\)) when the exact \(\sigma\) is supplied (Sec.~\ref{sec:AHO}). For
\(\gamma_\phi\), the algorithm with \(\sigma\) left free uses all eight loops and
does not yet yield converged values of either \(\sigma\) or \(\delta\)
(Sec.~\ref{sec:joint}).
\item \emph{Judgment.} The approximant built with the scenario's \(\sigma\) must
(a) reproduce the unused eight-loop coefficient and (b) predict an exponent
consistent with the scenario's exact \(\delta\). Criterion (a) certifies that
the approximant is a faithful representation of the series. A large error in (a)
is evidence against the scenario, but a small error is not by itself evidence
for it, since (a) can be satisfied for a range of \(\sigma\). Criterion (b) is the
discriminating one, but it is meaningful only for an approximant that passes
(a). A scenario is favored only if both criteria are satisfied.
\end{enumerate}
\end{revblock}

\section{Benchmark: the quartic anharmonic oscillator}
\label{sec:AHO}
\subsection{Approximant with the exact \texorpdfstring{\(\sigma\)}{sigma}}
\label{sec:AHOexact}

We first test the ideas above on a divergent series of the same type as that of
\(\gamma_{\phi}\), namely the ground-state energy of the quartic anharmonic
oscillator with Hamiltonian
\begin{equation}
H=\frac{p^{2}}{2}+\frac{1}{2}x^{2}+gx^{4}.  \label{phi4}
\end{equation}
\rev{This system is the \(\phi^4\) model in \(D=1\) dimension. Its large-order
behavior is governed by an instanton and it has no renormalons, which makes it
a natural benchmark for the four-dimensional case.} The weak-coupling
perturbation series for the ground-state energy is \cite{Bender-x4}
\begin{equation}
E_{0}=\frac{1}{2}+\frac{3}{4}g-\frac{21}{8}g^{2}+\frac{333}{16}g^{3}-\frac{
30885}{128}g^{4}+O(g^{5}).  \label{x4-pert}
\end{equation}
Its coefficients obey the large-order form
\(\gamma\, n!\,\sigma^{n} n^{\delta}\left( 1+O(1/n)\right)\), and in this case
all three parameters are known exactly: \(\sigma=-3\), \(\delta=-\tfrac{1}{2}\)
and \(\gamma=-\frac{\sqrt{6}}{\pi \Gamma\left(\frac{1}{2}\right)}\)
\rev{\cite{Bender-x4}}. We use the first five orders together with the known
\(\sigma\), and examine the effect of feeding \(\sigma\) into the hypergeometric
function on the prediction of the next order and on the predicted value of
\(\delta\).

Writing the series in Eq.~\eqref{x4-pert} as
\begin{equation}
E_{0}=\sum_{i=0}^{5}d_{i}g^{i}+O\left( g^{6}\right) ,
\end{equation}
we parametrize it by
\begin{equation}
E_{0}\simeq d_{0}+d_{1}g\,{}_{3}F_{1}(a_{1},a_{2},a_{3};b_{1};\sigma g).
\end{equation}
Expanding the hypergeometric function as
\begin{equation}
{}_{p}F_{q}(a_{1},\ldots,a_{p};b_{1},\ldots,b_{q};\sigma g)=\sum_{n=0}^{\infty}h_{n}g^{n},
\end{equation}
one has \cite{abohop}
\begin{eqnarray}
\frac{h_{n}}{h_{n-1}} &=&\frac{\prod_{i=1}^{p}\left( a_{i}+n-1\right)}{
n\prod_{j=1}^{q}\left( b_{j}+n-1\right)}\,\sigma ,  \notag \\
&=&\frac{\sum_{i=0}^{3}f_{i}n^{i}}{\sum_{j=1}^{2}e_{j}n^{j}},
\end{eqnarray}
so that, for \(p=3\), the linear system
\begin{equation}
\sum_{i=0}^{3}f_{i}n^{i}=\frac{h_{n}}{h_{n-1}}\sum_{j=1}^{2}e_{j}n^{j},
\qquad \rev{n=1,\ldots,4,}
\label{eq:linear-system}
\end{equation}
with \(e_{2}=1\) and \(\sigma=f_{3}\), determines the \(f_{i}\) and \(e_{j}\).
The numerator and denominator parameters are then the roots of
\begin{align}
\sum_{i=0}^{3}f_{i}\left( 1-A\right)^{i}& =0, \qquad a_{i}\ \text{the roots},  \notag \\
\sum_{j=1}^{2}e_{j}\left( 1-B\right)^{j-1}& =0, \qquad e_{p-1}=1,\quad b_{i}\ \text{the roots}.
\label{linear-a-b}
\end{align}

For the series in Eq.~\eqref{x4-pert}, the fifth-order approximant is
\begin{equation}
E_{0}\simeq \frac{1}{2}+\frac{3}{4}g\,{}_{3}F_{1}(a_{1},a_{2},a_{3};b_{1};-3g),
\label{Ehg3F1}
\end{equation}
where the exact \(\sigma=-3\) (instanton) has been used, and one finds \(a_{1}=-1.12275\),
\(a_{2}=0.420500\), \(a_{3}=2.77922\), and \(b_{1}=-1.12467\). As a first check of
the approximant, Eq.~\eqref{entire} gives \(E_{0}(0.5)=0.692069\), compared with
the exact (numerical) value \(0.696176\). Since for \(\gamma_{\phi}\) we shall
test the ability of the approximant to predict the next coefficient, we
re-expand Eq.~\eqref{Ehg3F1} and obtain
\begin{equation}
E_{0}=\frac{1}{2}+\frac{3}{4}g-\frac{21}{8}g^{2}+\frac{333}{16}g^{3}-\frac{
30885}{128}g^{4}+\frac{916731}{256}g^{5}-64430.9g^{6}.
\end{equation}
The exact coefficient is \(d_{6}=-\frac{65518401}{1024}=-63982.8\), so the
relative error is
\(\frac{\left\vert 64430.9-63982.8\right\vert }{63982.8} =0.70034\times 10^{-2}\),
i.e.\ below one percent.

The exact \(\delta\) of the original series is \(-\tfrac{1}{2}\); since the
parametrized series has one factor of \(g\) extracted, one unit must be added,
so the exact \(\delta\) for the parametrized series is \(\tfrac{1}{2}\). The
algorithm predicts \(\delta\) from Eq.~\eqref{constr}, giving \(\delta=1.20163\).
Compared with the convergence of \(\delta\) when the exact \(\sigma=-3\) is not
employed (see the last row of Table~1 in Ref.~\cite{entire}, where the predicted
\(\delta=2.764103\) is to be compared with the exact result \(-\frac{1}{2}\)),
this is a valuable improvement toward the exact value. The benchmark results are
summarized in Table~\ref{tab:AHO}.

\begin{table}[H]
\centering
\caption{\(\sigma\)-test for the quartic anharmonic oscillator, using the
five-loop series and the exact \(\sigma=-3\) as input.}
\label{tab:AHO}
\begin{tabular}{l c c c}
\toprule
Quantity & Predicted & Exact & Error \\
\midrule
\(E_0(g=0.5)\)  & 0.692069 & 0.696176 & 0.58994\%\\
\(d_6\)         & \(-64430.9\) & \(-63982.8\) & 0.70034\% \\
\(\delta\) (shifted) & 1.20163 & 0.5 & --- \\
\bottomrule
\end{tabular}
\end{table}

\begin{revblock}
\subsection{Control computation with wrong values of \texorpdfstring{\(\sigma\)}{sigma}}
\label{sec:AHOcontrol}

The accuracy of \(d_6\) at the exact \(\sigma\) is meaningful only if a wrong
input produces a noticeably worse prediction. We have therefore repeated the
construction for a range of wrong values of \(\sigma\), including the wrong sign.
The results are listed in Table~\ref{tab:AHOcontrol}.

\begin{table}[H]
\centering
\revcolor
\caption{Control computation for the anharmonic oscillator: the approximant of
Eq.~\eqref{Ehg3F1} built from \(d_0,\ldots,d_5\) with the input \(\sigma\) varied
from \(-3\) to \(+3\).
The exact values are \(\sigma=-3\), \(d_6=-63982.8\), and shifted
\(\delta=0.5\).}
\label{tab:AHOcontrol}
\begin{tabular}{c c c}
\toprule
input \(\sigma\) & error of \(d_6\) (\%) & predicted \(\delta\) \\
\midrule
\(-3\) (exact) & 0.70 & 1.20 \\
\(-2\) & 0.28 & 7.99 \\
\(-1\) & 0.46 & 41.6 \\
\(+1\) & 0.57 & \(-132.6\) \\
\(+2\) & 0.60 & \(-99.0\) \\
\(+3\) & 0.61 & \(-92.2\) \\
\bottomrule
\end{tabular}
\end{table}

The control computation shows why both steps and both criteria are needed. The
prediction of the next coefficient is essentially insensitive to \(\sigma\): every
value in Table~\ref{tab:AHOcontrol}, including those with the wrong sign,
reproduces \(d_6\) to better than \(1\%\), and some wrong values do better than
the exact one. With five orders as input, criterion (a) is therefore satisfied by
every input here and cannot by itself select \(\sigma\). The predicted exponent,
criterion (b), behaves quite differently. A wrong sign of \(\sigma\) produces
exponents of order \(-10^2\), and a wrong magnitude of \(\sigma\) with the
correct sign moves \(\delta_{\rm pred}\) away from the exact value, the more so
the farther \(\sigma\) is from \(-3\) (\(\delta=8.0\) for \(\sigma=-2\) and
\(41.6\) for \(\sigma=-1\)).  Only an input \(\sigma\) close to the exact one satisfies both
criteria: the next coefficient is reproduced (as it is for all inputs) and the
exponent is close to its exact value (as it is only near \(\sigma=-3\)). In the
oscillator, which has no renormalon, the two criteria taken together thus select
the correct large-order behavior. For \(\gamma_\phi\) we apply the same combined
judgment.
\end{revblock}

\section{Eight-loop tests of large-order asymptotics: relatively small
\texorpdfstring{\(N\)}{N}}
\label{sec:smallN}

The Lagrangian density of the \(O(N)\)-symmetric model is
\begin{equation}
\mathcal{L}=\frac{1}{2}\left( \partial \Phi \right)^{2}+\frac{m^{2}}{2}\Phi^{2}
+\frac{16\pi^{2}g}{4!}\left(\Phi^{2}\right)^{2},
\end{equation}
where \(\Phi=\left( \phi_{1},\phi_{2},\ldots,\phi_{N}\right)\) is an
\(N\)-component field. In minimal subtraction, the field anomalous dimension
\(\gamma_{\phi}\) has recently been obtained to eight loops by Schnetz
\rev{\cite{Schnetz2023}; see also Refs.~\cite{Schnetz2018,Schnetz-06-23}}. In four
dimensions it takes the form \rev{\cite{Schnetz2023}}
\begin{equation}
\gamma_{\phi}(g) = \frac{g^{2}}{36}(N+2) - \frac{g^{3}}{432}(N+2)[N+8] + \cdots + \mathcal{O}(g^{9}).
\end{equation}

In what follows we test the large-order asymptotic behavior of \(\gamma_{\phi}\)
for \(N=1,3,5,10,15\) and \(20\). In each case the seven-loop series is used,
together with either \(\sigma_{\rm inst}\) or \(\sigma_{\rm ren}\), to
parametrize a Gevrey-1 hypergeometric approximant of the form
\begin{equation}
\frac{\gamma_{\phi}}{g^{2}}\simeq c_{0}+c_{1}g\,
{}_{3}F_{1}(a_{1},a_{2},a_{3};b_{1};\sigma g),
\label{eq:param-gammaphi}
\end{equation}
which is then re-expanded in \(g\); the predicted eight-loop coefficient is
compared with the exact value, and the predicted exponent with the reference
values of Table~\ref{tab:exact}. \rev{Following Sec.~\ref{sec:AHOcontrol}, the two
predictions are judged together: an accurate eight-loop coefficient is required
of a favored scenario, and the predicted exponent must agree with the scenario's
exact value.} In
this section we consider the relatively small values \(N=1,3,5\); the larger
values \(N=10,15,20\) are treated in Sec.~\ref{sec:largeN}.

\subsection{The \texorpdfstring{\(N=1\)}{N=1} case}

For \(N=1\) the eight-loop series reads
\begin{equation*}
\frac{\gamma_{\phi}}{g^{2}}
=0.0833333-0.0625000g+0.338542g^{2}-1.92558g^{3}+14.3840g^{4}-124.159g^{5}+1171.88g^{6}.
\end{equation*}
Using the seven-loop coefficients and \(\sigma_{\rm inst}=-1\)
\cite{Kleinert-Borel} in Eq.~\eqref{eq:param-gammaphi}, the re-expanded
approximant gives
\begin{equation*}
\frac{\gamma_{\phi}\left( g\right)|^{\sigma_{\rm inst}}}{g^{2}}
=0.0833333-0.0625000g+0.338542g^{2}-1.92558g^{3}+14.3840g^{4}-124.159g^{5}+1205.77g^{6},
\end{equation*}
so that the relative error in the eight-loop coefficient is
\(\frac{\left\vert 1171.88-1205.77\right\vert }{1171.88}=2.8919\times 10^{-2}\).
For the renormalon, \(\sigma_{\rm ren}=\frac{N+8}{6}=\frac{3}{2}\)
\cite{Renormalon22}, and the re-expanded approximant is
\begin{equation*}
\frac{\gamma_{\phi}\left( g\right)|^{\sigma_{\rm ren}}}{g^{2}}
=0.0833333-0.0625000g+0.338542g^{2}-1.92558g^{3}+14.3840g^{4}-124.159g^{5}+1142.11g^{6},
\end{equation*}
with relative error
\(\frac{\left\vert 1171.88-1142.11\right\vert }{1171.88}=2.5404\times 10^{-2}\).

At this order the coefficient test does not discriminate between the two
scenarios. The predicted exponent does: the instanton parametrization gives
\(\delta=4.97451\), whereas the renormalon parametrization gives
\(\delta=-20.6151\). The reference values for the parametrized series
(Table~\ref{tab:exact}) are \(\delta_{\rm inst}=3+2+\frac{N}{2}=5.5\)
\cite{Kleinert-Borel} and
\(\delta_{\rm ren}=3-\frac{\left( 5N^2+116N+548\right)}{\left( N+8\right)^{2}}=-5.2593\).
Taken together, the eight-loop series for \(N=1\) favors the instanton
large-order behavior; see Table~\ref{tab:N1}.

\begin{table}[H]
\centering
\caption{\(\sigma\)-test for \(\gamma_\phi\) at \(N=1\). Exact eight-loop
coefficient: \(1171.88\).}
\label{tab:N1}
\begin{tabular}{l c c c c c}
\toprule
Scenario & \(\sigma\) & Predicted \(c_8\) & Error (\%) & Predicted \(\delta\) & Exact \(\delta\) \\
\midrule
Instanton  & \(-1\)  & 1205.77 & 2.8919 & 4.97451   & 5.5    \\
Renormalon & \(3/2\) & 1142.11 & 2.5404 & \(-20.6151\) & \(-5.2593\) \\
\bottomrule
\end{tabular}
\end{table}

\subsection{The \texorpdfstring{\(N=3\)}{N=3} case}

For \(N=3\) the eight-loop series is
\begin{equation}
\frac{\gamma_{\phi}\left( g\right) }{g^{2}}\simeq
0.138889-0.127315g+0.699267g^{2}-4.68924g^{3}+38.4364g^{4}-365.900g^{5}+3792.05g^{6}.
\end{equation}
Following the same recipe, the instanton-parametrized approximant re-expands to
\begin{equation}
\frac{\gamma_{\phi}\left( g\right)|^{\sigma_{\rm inst}}}{g^{2}}\simeq
0.138889-0.127315g+0.699267g^{2}-4.68924g^{3}+38.4364g^{4}-365.900g^{5}+3929.06g^{6},
\end{equation}
with error
\(\frac{\left\vert 3929.06-3792.05\right\vert }{3792.05}=3.6131\times 10^{-2}\),
while with \(\sigma_{\rm ren}=\frac{11}{6}\) one obtains
\begin{equation}
\frac{\gamma_{\phi}\left( g\right)|^{\sigma_{\rm ren}}}{g^{2}}\simeq
0.138889-0.127315g+0.699267g^{2}-4.68924g^{3}+38.4364g^{4}-365.900g^{5}+3885.43g^{6},
\end{equation}
with error
\(\frac{\left\vert 3885.43-3792.05\right\vert }{3792.05}=2.4625\times 10^{-2}\).
Again the coefficient test alone does not favor either scenario. The
predicted exponents, however, are \(\delta_{\rm inst}=6.79557\) and
\(\delta_{\rm ren}=-46.7968\), to be compared with the reference values
\(6.5\) and \(-4.7769\), respectively. On the basis of this second test the
series clearly favors \(\sigma_{\rm inst}\) over \(\sigma_{\rm ren}\)
(Table~\ref{tab:N3}).

\begin{table}[H]
\centering
\caption{\(\sigma\)-test for \(\gamma_\phi\) at \(N=3\). Exact eight-loop
coefficient: \(3792.05\).}
\label{tab:N3}
\begin{tabular}{l c c c c c}
\toprule
Scenario & \(\sigma\) & Predicted \(c_8\) & Error (\%) & Predicted \(\delta\) & Exact \(\delta\) \\
\midrule
Instanton  & \(-1\)   & 3929.06 & 3.6131 & 6.79557     & 6.5    \\
Renormalon & \(11/6\) & 3885.43 & 2.4625 & \(-46.7968\) & \(-4.7769\) \\
\bottomrule
\end{tabular}
\end{table}

\subsection{The \texorpdfstring{\(N=5\)}{N=5} case}

For the moderately large value \(N=5\) the eight-loop series is
\begin{equation}
\frac{\gamma_{\phi}\left( g\right) }{g^{2}}\simeq
0.194444 - 0.210648 g + 1.11400 g^2 - 8.77905 g^3 + 77.5264 g^4 -799.547 g^5 + 8946.35 g^6.
\end{equation}
The seven-loop instanton parametrization re-expands to
\begin{equation}
\frac{\gamma_{\phi}\left( g\right)|^{\sigma_{\rm inst}}}{g^{2}}\simeq
0.194444 - 0.210648 g + 1.11400 g^2 - 8.77905 g^3 + 77.5264 g^4 -799.547 g^5 + 9193.29 g^6,
\end{equation}
predicting the eight-loop coefficient with error
\(\frac{\left\vert 9193.29-8946.35\right\vert }{8946.35}=2.7602\times 10^{-2}\),
whereas the renormalon parametrization gives
\begin{equation}
\frac{\gamma_{\phi}\left( g\right)|^{\sigma_{\rm ren}}}{g^{2}}\simeq
\rev{0.194444 - 0.210648 g + 1.11400 g^2 - 8.77905 g^3 + 77.5264 g^4 -799.547 g^5}
+11113.3g^{6},
\end{equation}
with error
\(\frac{\left\vert 11113.3-8946.35\right\vert }{8946.35}=0.24222 \).
\rev{The difference between the two coefficient errors is now large. The exponent
test points in the same direction:}
the instanton parametrization predicts \(\delta_{\rm inst}=7.13702\) against
the reference value \(7.5\), while the renormalon parametrization predicts
\(\delta_{\rm ren}=-2.90269\) against \(-4.4142\) (Table~\ref{tab:N5}).
\rev{At \(N=5\) the renormalon exponent is, however, much closer to its reference
value than for the other values of \(N\); this case is examined further in
Sec.~\ref{sec:joint}.}

\begin{table}[H]
\centering
\caption{\(\sigma\)-test for \(\gamma_\phi\) at \(N=5\). Exact eight-loop
coefficient: \(8946.35\).}
\label{tab:N5}
\begin{tabular}{l c c c c c}
\toprule
Scenario & \(\sigma\) & Predicted \(c_8\) & Error (\%) & Predicted \(\delta\) & Exact \(\delta\) \\
\midrule
Instanton  & \(-1\)   & 9193.29 & 2.7602 & 7.13702     & 7.5     \\
Renormalon & \(13/6\) & 11113.3 & 24.222  & \(-2.90269\) & \(-4.4142\) \\
\bottomrule
\end{tabular}
\end{table}

\section{Eight-loop tests of large-order asymptotics: larger
\texorpdfstring{\(N\)}{N}}
\label{sec:largeN}

For \(N=5\) the coefficient test already separates the
\(\sigma_{\rm inst}\) and \(\sigma_{\rm ren}\) parametrizations, in favor of
the instanton. We extend the test in this section to \(N=10\), \(15\) and \(20\).

\subsection{The \texorpdfstring{\(N=10\)}{N=10} case}

For \(N=10\) the eight-loop series is
\begin{equation}
\frac{\gamma_{\phi}\left( g\right) }{g^{2}}\simeq
0.333333-0.500000g+2.08333g^{2}-25.2267g^{3}+257.408g^{4}-3095.56g^{5}+40212.5g^{6}.
\end{equation}
Parametrizing Eq.~\eqref{eq:param-gammaphi} with \(\sigma_{\rm inst}=-1\) and
the seven-loop coefficients, we obtain
\begin{equation}
\frac{\gamma_{\phi}\left( g\right)|^{\sigma_{\rm inst}}}{g^{2}}\simeq
0.333333-0.500000g+2.08333g^{2}-25.2267g^{3}+257.408g^{4}-3095.56g^{5}+41650.1g^{6},
\end{equation}
i.e.\ a relative error of
\(\frac{\left\vert 41650.1-40212.5\right\vert }{40212.5} =0.03575 \)
in the eight-loop coefficient. For \(\sigma_{\rm ren}=\frac{18}{6}=3\) we obtain
\begin{equation}
\frac{\gamma_{\phi}\left( g\right)|^{\sigma_{\rm ren}}}{g^{2}}\simeq
0.333333-0.500000g+2.08333g^{2}-25.2267g^{3}+257.408g^{4}-3095.56g^{5}+31010.0g^{6},
\end{equation}
with error
\(\frac{\left\vert 31010.0-40212.5\right\vert }{40212.5} =0.22885 \).

Finally, the \(\sigma_{\rm inst}\) parametrization predicts \(\delta=10.0256\)
against the reference value \(10\), while the \(\sigma_{\rm ren}\)
parametrization predicts \(\delta=-9.77562\) against \(-3.8148\). Both tests
therefore favor the instanton large-order behavior (Table~\ref{tab:N10}).

\begin{table}[H]
\centering
\caption{\(\sigma\)-test for \(\gamma_\phi\) at \(N=10\). Exact eight-loop
coefficient: \(40212.5\).}
\label{tab:N10}
\begin{tabular}{l c c c c c}
\toprule
Scenario & \(\sigma\) & Predicted \(c_8\) & Error (\%) & Predicted \(\delta\) & Exact \(\delta\) \\
\midrule
Instanton  & \(-1\) & 41650.1 & 3.575  & 10.0256      & 10     \\
Renormalon & \(3\)  & 31010.0 & 22.885 & \(-9.77562\) & \(-3.8148\) \\
\bottomrule
\end{tabular}
\end{table}

\subsection{The \texorpdfstring{\(N=15\)}{N=15} case}

To check that this preference persists with increasing \(N\), we consider
\(N=15\), for which the eight-loop series is
\begin{equation}
\frac{\gamma_{\phi}\left( g\right) }{g^{2}}\simeq
0.472222-0.905093g+2.37751g^{2}-50.5249g^{3}+579.570g^{4}-7800.69g^{5}+113619g^{6}.
\end{equation}
Using the seven-loop series together with \(\sigma_{\rm inst}\), the re-expanded
approximant is
\begin{equation}
\frac{\gamma_{\phi}\left( g\right)|^{\sigma_{\rm inst}}}{g^{2}}\simeq
0.472222-0.905093g+2.37751g^{2}-50.5249g^{3}+579.570g^{4}-7800.69g^{5}+117232g^{6},
\end{equation}
with relative error
\(\frac{\left\vert 117232-113619\right\vert }{113619}  =0.031799 \)
in the eight-loop coefficient. With \(\sigma_{\rm ren}=\frac{23}{6}\) we obtain
\begin{equation}
\frac{\gamma_{\phi}\left( g\right)|^{\sigma_{\rm ren}}}{g^{2}}\simeq
0.472222-0.905093g+2.37751g^{2}-50.5249g^{3}+579.570g^{4}-7800.69g^{5}+93936.0g^{6},
\end{equation}
with error
\(\frac{\left\vert 93936.0-113619\right\vert }{113619}=0.17324 \).
The instanton parametrization predicts \(\delta=12.2451\) against the reference
value \(12.5\), while the renormalon parametrization predicts
\(\delta=-10.2396\) against \(-3.4518\) (Table~\ref{tab:N15}).

\begin{table}[H]
\centering
\caption{\(\sigma\)-test for \(\gamma_\phi\) at \(N=15\). Exact eight-loop
coefficient: \(113619\).}
\label{tab:N15}
\begin{tabular}{l c c c c c}
\toprule
Scenario & \(\sigma\) & Predicted \(c_8\) & Error (\%) & Predicted \(\delta\) & Exact \(\delta\) \\
\midrule
Instanton  & \(-1\)   & 117232  & 3.1799 & 12.2451      & 12.5   \\
Renormalon & \(23/6\) & 93936.0 & 17.324 & \(-10.2396\) & \(-3.4518\) \\
\bottomrule
\end{tabular}
\end{table}

\subsection{The \texorpdfstring{\(N=20\)}{N=20} case}

For \(N=20\) the eight-loop series is
\begin{equation}
\frac{\gamma_{\phi}\left( g\right) }{g^{2}}\simeq
0.611111-1.42593g+1.27315g^{2}-83.3471g^{3}+1066.86g^{4}-15636.4g^{5}+250291g^{6}.
\end{equation}
The seven-loop parametrization predicts the eight-loop coefficient as
\(253894\), i.e.\ within
\(\frac{\left\vert 253894-250291\right\vert }{250291}\times 100\%=1.4395\%\),
for \(\sigma_{\rm inst}=-1\), and as \(205285\), i.e.\ within
\(\frac{\left\vert 205285-250291\right\vert }{250291}\times 100\%=17.981\%\),
for \(\sigma_{\rm ren}=\frac{14}{3}\). The two parametrizations predict
\(\delta=13.7301\) and \(\delta=-10.1269\), respectively, against the
reference values \(\delta_{\rm inst}=15\) and \(\delta_{\rm ren}=\rev{-3.2092}\)
(Table~\ref{tab:N20}).

\begin{table}[H]
\centering
\caption{\(\sigma\)-test for \(\gamma_\phi\) at \(N=20\). Exact eight-loop
coefficient: \(250291\).}
\label{tab:N20}
\begin{tabular}{l c c c c c}
\toprule
Scenario & \(\sigma\) & Predicted \(c_8\) & Error (\%) & Predicted \(\delta\) & Exact \(\delta\) \\
\midrule
Instanton  & \(-1\)   & 253894 & 1.4395 & 13.7301    & 15     \\
Renormalon & \(14/3\) & 205285 & 17.981 & \(-10.1269\) & \rev{\(-3.2092\)} \\
\bottomrule
\end{tabular}
\end{table}

\begin{revblock}
\section{Sign versus magnitude of \texorpdfstring{\(\sigma\)}{sigma}, and the
joint \texorpdfstring{\((\sigma,\delta)\)}{(sigma,delta)} test}
\label{sec:joint}

The renormalon and instanton values of \(\sigma\) differ both in sign and in
magnitude. Since the alternation of the coefficients of \(\gamma_\phi\) already
indicates \(\sigma<0\), the comparison of Secs.~\ref{sec:smallN}
and~\ref{sec:largeN} would add little if it only reflected the sign. Moreover,
by Eq.~\eqref{eq:delta-sigma}, the large negative exponents obtained with
positive \(\sigma\) follow to a large extent from the sign of the coefficient
\(A\), which is itself tied to the alternation of the series. We therefore
separate the two roles in two ways.

\subsection{Correct sign, wrong magnitude}

Table~\ref{tab:signmag} compares four input values for each \(N\): the instanton
value \(\sigma=-1\); the renormalon value \(\sigma=+(N+8)/6\); the renormalon
magnitude with the instanton sign, \(\sigma=-(N+8)/6\); and the instanton
magnitude with the renormalon sign, \(\sigma=+1\).

\begin{table*}[h]
\centering
\revcolor
\caption{Separation of the sign and the magnitude of \(\sigma\) for
\(\gamma_\phi\). For each input \(\sigma\) the relative error of the predicted
eight-loop coefficient (in \%) and the predicted exponent are given. The
instanton reference exponent is \(\delta_{\rm inst}=5+N/2\).}
\label{tab:signmag}
\begin{tabular}{c c | c c | c c | c c | c c}
\toprule
& & \multicolumn{2}{c|}{\(\sigma=-1\)} & \multicolumn{2}{c|}{\(\sigma=+(N+8)/6\)} &
\multicolumn{2}{c|}{\(\sigma=-(N+8)/6\)} & \multicolumn{2}{c}{\(\sigma=+1\)} \\
\(N\) & \(\delta_{\rm inst}\) & err.\ & \(\delta_{\rm pred}\) & err.\ & \(\delta_{\rm pred}\) &
err.\ & \(\delta_{\rm pred}\) & err.\ & \(\delta_{\rm pred}\) \\
\midrule
1  & 5.5  & 2.9 & 4.97  & 2.5  & \(-20.6\) & 5.9  & 1.36    & 2.0  & \(-24.2\) \\
3  & 6.5  & 3.6 & 6.80  & 2.5  & \(-46.8\) & 23.1 & 1.57    & 2.6  & \(-52.0\) \\
5  & 7.5  & 2.8 & 7.14  & 24.2 & \(-2.90\) & 7.7  & \(-0.27\) & 47.9 & \(-10.3\) \\
10 & 10   & 3.6 & 10.03 & 22.9 & \(-9.78\) & 11.1 & \(-0.84\) & 7.0  & \(-20.6\) \\
15 & 12.5 & 3.2 & 12.25 & 17.3 & \(-10.2\) & 12.9 & \(-1.39\) & 4.6  & \(-23.9\) \\
20 & 15   & 1.4 & 13.73 & 18.0 & \(-10.1\) & 13.3 & \(-1.84\) & 5.3  & \(-25.7\) \\
\bottomrule
\end{tabular}
\end{table*}

The comparison between the first and third pairs of columns isolates the
magnitude of \(\sigma\) at fixed (correct) sign. Replacing \(|\sigma|=1\) by the
renormalon magnitude \((N+8)/6\) increases the coefficient error for every \(N\),
from \(1.4\)--\(3.6\%\) to \(5.9\)--\(23\%\). More importantly, it moves the
predicted exponent to values between \(-1.8\) and \(1.6\), far from
\(\delta_{\rm inst}=5+N/2\), and it removes all \(N\)-dependence of the kind
predicted by the instanton. With \(|\sigma|=1\), by contrast, the predicted
exponents follow \(5+N/2\) closely, increasing from \(5.0\) at \(N=1\) to
\(13.7\) at \(N=20\). This information is not contained in the alternation of
the series.

\subsection{Joint test: the value of \texorpdfstring{\(\sigma\)}{sigma} required by
the exponent}

A sharper formulation reverses the logic of the \(\sigma\)-test. Instead of
fixing \(\sigma\) and comparing the predicted exponent with its reference value,
we fix the exponent and ask which \(\sigma\) the data require. By
Eq.~\eqref{eq:delta-sigma}, the condition
\(\delta_{\rm pred}(\sigma)=\delta_{\rm ref}\) is a quadratic equation for
\(\sigma\), with no adjustable parameter. If the data carry the instanton imprint,
the root \(\sigma_\delta\) obtained with \(\delta_{\rm ref}=5+N/2\) should lie
close to \(-1\). If they carry the renormalon imprint, the condition
\(\delta_{\rm pred}(\sigma)=\delta_{\rm ren}\) should have a root close to
\((N+8)/6\).

The results for \(\gamma_\phi\) are collected in Table~\ref{tab:joint}
and illustrated in Fig.~\ref{fig:scan}(a).

\begin{table}[H]
\centering
\revcolor
\caption{Joint \((\sigma,\delta)\) test. \(\sigma_\delta^{\rm inst}\): value of
\(\sigma\) at which the predicted exponent equals \(\delta_{\rm inst}=5+N/2\)
(root closest to \(-1\)). \(\sigma_\delta^{\rm ren}\): positive value of
\(\sigma\) at which it equals \(\delta_{\rm ren}\) of Table~\ref{tab:exact}
(positive root of the quadratic equation). ``none'': the quadratic
equation has no positive root.}
\label{tab:joint}
\begin{tabular}{c c c c c}
\toprule
\(N\) & \(\sigma_\delta^{\rm inst}\) & \(\sigma_{\rm inst}\) &
\(\sigma_\delta^{\rm ren}\) & \(\sigma_{\rm ren}\) \\
\midrule
1  & \(-0.957\) & \(-1\) & none & 1.500 \\
3  & \(-1.018\) & \(-1\) & none & 1.833 \\
5  & \(-0.970\) & \(-1\) & 1.81 & 2.167 \\
10 & \(-1.002\) & \(-1\) & 11.7 & 3.000 \\
15 & \(-0.986\) & \(-1\) & 36.0 & 3.833 \\
20 & \(-0.940\) & \(-1\) & 162 & 4.667 \\
\bottomrule
\end{tabular}
\end{table}

\begin{figure*}[t]
\centering
\revcolor
\includegraphics[width=\textwidth]{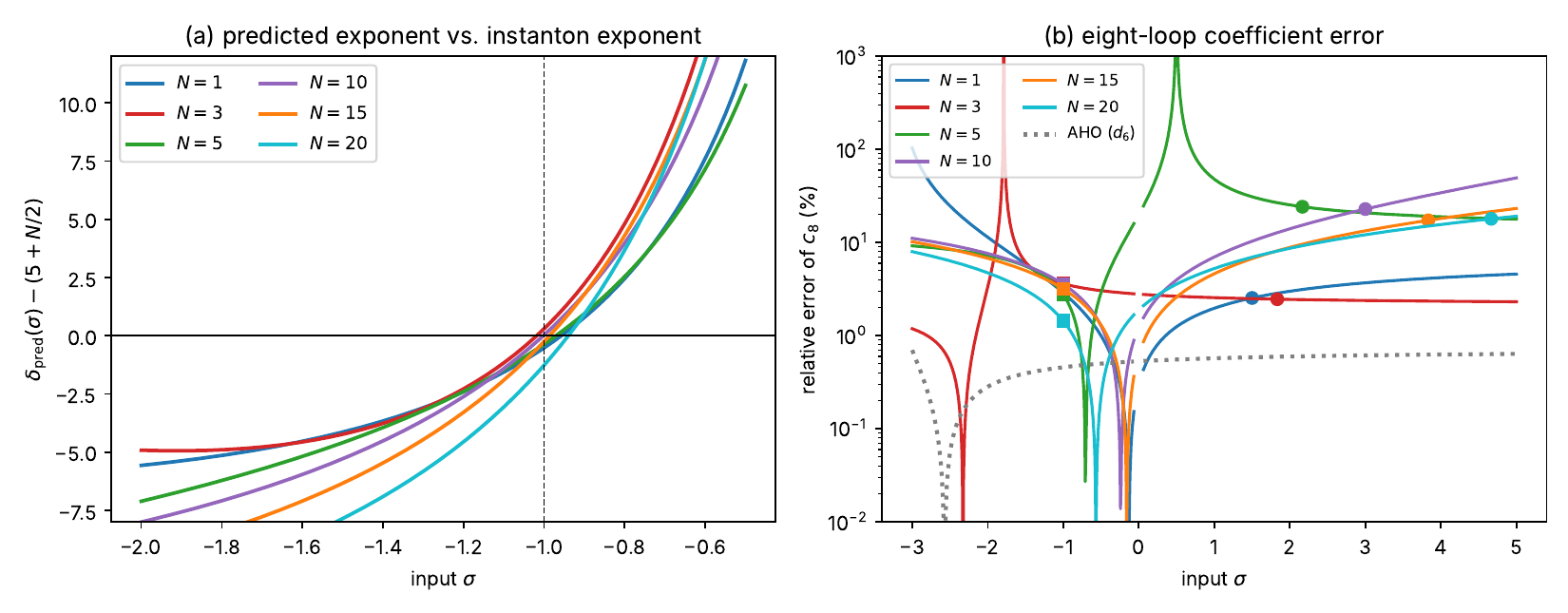}
\caption{(a) Difference between the exponent predicted by the seven-loop
approximant, Eq.~\eqref{eq:delta-sigma}, and the instanton exponent \(5+N/2\), as
a function of the input \(\sigma\). For every \(N\) the curve crosses zero within
about \(6\%\) of \(\sigma_{\rm inst}=-1\) (dashed line). (b) Relative error of the
predicted eight-loop coefficient as a function of the input \(\sigma\); squares:
\(\sigma_{\rm inst}=-1\), circles: \(\sigma_{\rm ren}=(N+8)/6\). The dotted gray
curve is the corresponding error of \(d_6\) for the anharmonic oscillator, which
is nearly independent of \(\sigma\) and illustrates the limited discriminating
power of the coefficient test alone. Because the parameters of the linear system
are affine in \(\sigma\), the predicted coefficient is a ratio of two linear
functions of \(\sigma\). The upward spikes are its pole, where the denominator
parameter reaches \(b_1=-4\), so that \((b_1)_5=0\) (e.g.\ \(\sigma=-3.20\) for
\(N=1\), \(-1.79\) for \(N=3\), \(+0.50\) for \(N=5\), and \(-3.43\) for the
oscillator). The downward spikes are zeros of the error, where the prediction
coincides with the exact coefficient (the values listed at the end of
Sec.~\ref{sec:joint}).}
\label{fig:scan}
\end{figure*}

The instanton values pass the joint test for every \(N\): the value of
\(\sigma\) at which the predicted exponent equals \(5+N/2\) lies between
\(-0.94\) and \(-1.02\), i.e.\ within about \(6\%\) of \(\sigma_{\rm inst}=-1\). This is
a nontrivial statement. The required \(\sigma\) is not imposed but determined
by the data. The instanton reference exponent \(5+N/2\) varies by almost a
factor of three over the range of \(N\) considered (from \(5.5\) to \(15\)), and
the \(N\)-dependence of the low-order coefficients is substantial (the ratio
\(c_5/c_4\) changes from \(-8.6\) at \(N=1\) to \(-14.7\) at \(N=20\)); the root
nevertheless stays at \(-1\). The quadratic equation
\eqref{eq:joint-quadratic} with \(\delta_{\rm ref}=5+N/2\) has a second root in
each case (\(-9.1\) and \(-3.6\) for \(N=1\) and \(3\); \(+6.7\), \(+64\),
\(+231\) and \(+1174\) for \(N=5,10,15,20\)), which is far from both
\(\sigma_{\rm inst}\) and \(\sigma_{\rm ren}\) and is therefore discarded.

The renormalon values do not pass the test. For \(N=1\) and \(3\) there is no
positive \(\sigma\) at which the predicted exponent reaches the renormalon
reference value. For \(N=10\), \(15\) and
\(20\) a positive solution exists, but it lies at \(\sigma\simeq 12\), \(36\) and
\(160\), i.e.\ 4, 9 and 35 times \(\sigma_{\rm ren}\), and it moves further away
from \(\sigma_{\rm ren}\) as \(N\) grows, whereas the instanton solution stays at
\(-1\). The exception is \(N=5\), where the required value
\(1.81\) is close to \(\sigma_{\rm ren}=13/6\); here the joint test alone
cannot exclude a renormalon-type parametrization, and the preference for the
instanton at \(N=5\) rests on the large difference between the coefficient errors
(\(2.8\%\) versus \(24\%\)). The outcome does not depend on the integer offset of Step~4 of
Sec.~\ref{sec:exact}: shifting the renormalon reference exponent by \(\pm1\)
moves the positive root to \(2.0\)--\(1.6\) for \(N=5\) and to \(15\)--\(9\),
\(47\)--\(26\) and \(216\)--\(109\) for \(N=10,15,20\), and still leaves no
positive root for \(N=1,3\). The same holds if the exponent of
Ref.~\cite{Renormalon22}, Eq.~\eqref{renormalon-ab}, is used as the renormalon
reference: there is again no positive root for \(N=1,3\), and the positive root
lies at \(4.9\), \(39\), \(121\) and \(536\) for \(N=5,10,15,20\).

Finally, the results also illustrate step (i) of the test. If \(\sigma\) is not
supplied but predicted by the algorithm, all eight loops are needed to fix the
\({}_3F_1\) parameters, and the predicted value is \(\sigma=-0.13\), \(-2.33\),
\(-0.71\), \(-0.24\), \(-0.16\) and \(-0.57\) for \(N=1,3,5,10,15,20\),
respectively. By construction, this is the value of \(\sigma\) at which the
eight-loop coefficient is reproduced exactly. At these values the predicted
exponent is between \(2\) and \(108\). The negative sign is correctly
identified, but neither \(\sigma\) nor \(\delta\) has converged at eight loops.
This is precisely why the exact \(\sigma\) is supplied in step (i). It also shows
again that reproducing the next coefficient, criterion (a), must be supplemented
by the exponent, criterion (b).

\end{revblock}

\begin{revblock}
\section{Relation to high-order primitive data and the approach to
asymptotics}
\label{sec:asymptotics}

The studies of primitive graphs to 17--18 loops
\cite{Balduf2023,BaldufShaban2024,BaldufThurigen2024,BorinskyFavorito2025}
show that the asymptotic regime of \(\phi^4\) theory is approached slowly. In
the zero-dimensional \(O(N)\) model the leading growth rate becomes visible only
above about 25 loops, and lower orders suggest a wrong asymptotics. In four
dimensions the growth rate of the primitive beta function differs from its
asymptotic value even at 17 loops, while its \(N\)-dependence is close to the
asymptotic form already at about ten loops \cite{BaldufThurigen2024}. These
authors also stress that agreement of particular quantities with the asymptotics
below ten loops must not be mistaken as evidence that the asymptotic regime has
been reached. This bears directly on how the present results should be read.

First, the \(\sigma\)-test does not extract the large-order parameters from the
data in the asymptotic sense. It asks whether a Gevrey-1 hypergeometric
function with prescribed large-order parameters is consistent with the known
coefficients, i.e.\ which large-order structure the available orders are
compatible with. The result of Sec.~\ref{sec:joint} is that the seven- and
eight-loop coefficients of \(\gamma_\phi\) are compatible, for all \(N\)
considered, with the instanton pair \((\sigma,\delta)=(-1,5+N/2)\), including
its \(N\)-dependence. This is in line with the observation of
Ref.~\cite{BaldufThurigen2024} that the \(N\)-dependence approaches its
asymptotic form early. 

Second, the result cannot exclude a renormalon that becomes dominant only at
higher orders. Since \(\sigma_{\rm ren}=(N+8)/6>1=|\sigma_{\rm inst}|\) for all
\(N\), a renormalon singularity with nonzero residue would lie closer to the
origin of the Borel plane and would eventually dominate. However, its exponent is
much smaller than the instanton exponent,
\(\delta_{\rm inst}-\delta_{\rm ren}\simeq 11\)--\(18\) for \(1\le N\le 20\).
For equal normalizations the two contributions,
\(\sigma_{\rm ren}^k k^{\delta_{\rm ren}}\) and \(k^{\delta_{\rm inst}}\), become
comparable only at
\begin{equation}
k^\ast\ln\sigma_{\rm ren}=(\delta_{\rm inst}-\delta_{\rm ren})\ln k^\ast ,
\end{equation}
which gives \(k^\ast\approx 129,\ 82,\ 64,\ 49,\ 45,\ 45\) for
\(N=1,3,5,10,15,20\). A renormalon of comparable strength would therefore be
invisible at eight loops, and a different normalization ratio \(R\) shifts
\(k^\ast\) only logarithmically, by about \(\ln R/\ln\sigma_{\rm ren}\). The
present data are thus consistent both with the absence of renormalons in the
\(\overline{\rm MS}\) RG functions \cite{BergereDavid1984,McKane2019,Renormalon22}
and with a renormalon whose effect is delayed to very high orders; they are not
consistent with a renormalon that already dominates the low orders.
\end{revblock}

\section{Summary and conclusions}
\label{sec:conclusions}

\begin{revblock}
We have proposed a \(\sigma\)-test, based on the hypergeometric resummation
algorithm developed in Refs.~\cite{eps7,abohop,abo-precize,Abo-large}, to
decide whether the eight-loop series for the field anomalous dimension
\(\gamma_\phi\) of the \(O(N)\)-symmetric \(\phi^4\) theory carries instanton or
renormalon large-order structure. The test uses the value of \(\sigma\) of each
scenario as input to a seven-loop \({}_3F_1\) approximant and examines its
prediction of the unused eight-loop coefficient and of the exponent \(\delta\).
All results are collected in Table~\ref{tab:summary}.

\begin{table*}[h]
\centering
\revcolor
\caption{Summary of the \(\sigma\)-test for \(\gamma_\phi\). Errors refer to the
prediction of the eight-loop coefficient from the seven-loop series with
\(\sigma\) fixed; exponents are for the parametrized series (see
Table~\ref{tab:exact}); \(\sigma_\delta^{\rm inst}\) is the value of \(\sigma\)
required by the instanton exponent (Table~\ref{tab:joint}).}
\label{tab:summary}
\begin{tabular}{c c c c c c c c c}
\toprule
& \multicolumn{2}{c}{Error of \(c_8\) (\%)} &
\multicolumn{2}{c}{\(\delta_{\rm inst}\)} &
\multicolumn{2}{c}{\(\delta_{\rm ren}\)} & & \\
\cmidrule(lr){2-3}\cmidrule(lr){4-5}\cmidrule(lr){6-7}
\(N\) & inst. & ren. & pred. & exact & pred. & exact &
\(\sigma_\delta^{\rm inst}\) & Favored \\
\midrule
1  & 2.8919 & 2.5404 & 4.97451 & 5.5  & \(-20.6151\) & \(-5.2593\) & \(-0.957\) & instanton \\
3  & 3.6131 & 2.4625 & 6.79557 & 6.5  & \(-46.7968\) & \(-4.7769\) & \(-1.018\) & instanton \\
5  & 2.7602 & 24.222 & 7.13702 & 7.5  & \(-2.90269\) & \(-4.4142\) & \(-0.970\) & instanton\(^{a}\) \\
10 & 3.575  & 22.885 & 10.0256 & 10   & \(-9.77562\) & \(-3.8148\) & \(-1.002\) & instanton \\
15 & 3.1799 & 17.324 & 12.2451 & 12.5 & \(-10.2396\) & \(-3.4518\) & \(-0.986\) & instanton \\
20 & 1.4395 & 17.981 & 13.7301 & 15   & \(-10.1269\) & \(-3.2092\) & \(-0.940\) & instanton \\
\bottomrule
\end{tabular}
\\[2pt]
\raggedright
\footnotesize{\(^{a}\)At \(N=5\) the joint test alone does not exclude the
renormalon parametrization (Table~\ref{tab:joint}); it is disfavored because it
misses the eight-loop coefficient by \(24\%\).}
\end{table*}

The test has two steps. Supplying the exact \(\sigma\) accelerates the convergence
of the predicted exponent, and the scenario is then judged by the eight-loop
coefficient and the exponent taken together. The control computations on the
anharmonic oscillator, the \(\phi^4\) model in \(D=1\), show that the next
coefficient is reproduced for a wide range of \(\sigma\), whereas the predicted
exponent is highly sensitive to it; only near the exact \(\sigma\) are both
criteria met. For \(\gamma_\phi\), the instanton parametrization meets both
criteria for every \(N\): the eight-loop coefficient is reproduced to
\(1.4\)--\(3.6\%\) and the predicted exponent follows \(5+N/2\). The renormalon
parametrization fails at least one criterion for every \(N\). For \(N=1\) and
\(3\) it reproduces the coefficient but gives exponents far from the reference
value. For \(N\ge10\) it fails both. For \(N=5\) its exponent is
closer to the reference value, but the coefficient is missed by \(24\%\).

The complementary joint \((\sigma,\delta)\) test, in which the value of
\(\sigma\) required by a given exponent is determined from the data, confirms this
picture. For every \(N\) between 1 and 20, the
value of \(\sigma\) at which the approximant reproduces the instanton exponent
\(5+N/2\) lies within about \(6\%\) of \(\sigma_{\rm inst}=-1\). Keeping the negative
sign but using the renormalon magnitude \((N+8)/6\) destroys this agreement, so
the test probes the magnitude of \(\sigma\) together with the exponent and its
\(N\)-dependence, and not only the sign visible in the alternation of the
series. With the renormalon parameters the required \(\sigma\) either does not
exist (\(N=1,3\)) or lies far above \(\sigma_{\rm ren}\) (\(N=10,15,20\)); the
exception is \(N=5\), where the joint test alone is inconclusive and the decision rests on the
eight-loop coefficient.

We conclude that the low-order coefficients of the
\(\overline{\mathrm{MS}}\) field anomalous dimension carry the instanton
large-order structure and show no visible renormalon component, in agreement with
the conclusions drawn for the beta function in
Refs.~\cite{Borel-6L,Schnetz2017,Renormalon22}. In view of the slow approach to
the asymptotic regime found in the high-loop studies of primitive graphs
\cite{Balduf2023,BaldufShaban2024,BaldufThurigen2024,BorinskyFavorito2025}, and
because a renormalon of comparable strength would only become visible at
\(k\gtrsim 45\)--\(130\) loops, this does not by itself establish that the
renormalon is absent asymptotically. It does show that the available orders are
governed by the instanton.
\end{revblock}

\section*{Declaration of generative AI and AI-assisted technologies in the writing process}

During the preparation of this work, the author used generative AI tools to
polish the wording of the manuscript. The author also used Claude (Anthropic) to
assist in deriving the modification of the renormalon large-order behavior due to
the \(\phi^6\) insertion.
After using these tools, the author reviewed and edited the content as needed and
takes full responsibility for the content of the publication.

\bibliographystyle{unsrtnat}
\bibliography{Renormalon-checked,renormalon-additions-checked}

\end{document}